\documentclass[preprint,amsmath,amssymb]{elsarticle}
\usepackage{graphicx}

\begin{document}

\title{Modelling the electronic structure and magnetic properties of LiFeAs and FeSe using hybrid-exchange density functional theory}

\author{Wei Wu}
\ead{w.wu@imperial.ac.uk}
\address{Department of Materials and London Centre for Nanotechnology, Imperial College London, South Kensington Campus, London, SW7 2AZ, U.K.}

\begin{abstract}
The electronic structure and magnetic properties of LiFeAs and FeSe have been studied using hybrid exchange density functional theory. The total energies for a unit cell in LiFeAs and FeSe with different spin states including non-magnetic and spin-2 are calculated. The spin-2 configuration has the lower energy for both LiFeAs and FeSe. The computed anti-ferromagnetic exchange interactions between spins on the nearest (next nearest) neighbouring Fe atoms in LiFeAs and FeSe are approximately $14$ ($17$) meV and $6$ ($13$) meV respectively. The total energies of the checkerboard and stripe-type anti-ferromagnetic ordering for LiFeAs and FeSe are compared, yielding that for LiFeAs the checkerboard is lower whereas for FeSe the stripe-type is lower. However, owing to the fact that the exchange interaction of the next nearest neighbour is larger than that of the nearest one, which means that the collinear ordering might be the ground state. These results are in agreement with previous theoretical calculations and experiments. Especially the calculations for LiFeAs indicate a co-existence of conducting $d$-bands at the Fermi surface and $d$-orbital magnetism far below the Fermi surface. The theoretical results presented here might be useful for the experimentalists working on the electronic structure and magnetism of iron-based superconductors.  

\smallskip

\noindent Author keywords: A. superconducting materials; D. Electronic structure; D. Magnetism

\end{abstract}

\maketitle

\section{Introduction}
\begin{figure}[htbp]
\begin{tabular}{c}
\includegraphics[scale=0.4, clip=true]{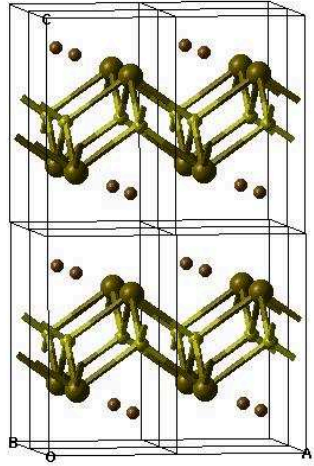}\\
(a)\\
\includegraphics[scale=0.3, clip=ture]{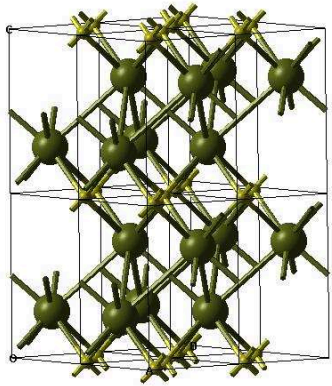}\\
(b)\\
\end{tabular}
\caption{(Colour on line.) The structures of a $2a\times 2b\times 2c$ super-cell of LiFeAs (a) and FeSe (b) are shown. In (a) small brown balls are used to label Li atoms, large brown for As, and small yellow for Fe. In (b) large green balls are used to label Se, and small yellow for Fe.}\label{fig:crystalstructure}
\end{figure}

Recently iron-based superconductors (FeSCs) have attracted much attention following the discovery of the high-$T_c$ superconductivity (HTSC) in LaFeAs(O,F) \cite{kamihara2008}.  The FeSCs that have received extensive experimental and theoretical studies include ReFeAsO (Re=La or rare earth) (1111-type) \cite{kamihara2008, ren2008}, A$\mathrm{Fe}_2\mathrm{As}_2$ (A=Ba, Sr, Ca, etc) (122-type) \cite{rotter2008}, LiFeAs (Fig.\ref{fig:crystalstructure}a) or NaFeAs (111-type) \cite{tapp2008}, and FeSe (11-type) \cite{hsu2008} (Fig.\ref{fig:crystalstructure}b). The family of FeSCs as a whole has the second highest $T_c$, and are more tunable than cuprate-based superconductors (CuSCs) as one can control transition temperature not only by doping but also by applying pressure on parent stoichiometric compounds. 

The common structural properties among these FeSCs is that Fe atoms form two-dimensional layers, in which spins on Fe atoms interact with each other via a super-exchange mediated by non-metallic species \cite{abrahams2011}. This can be naturally modelled by using a two-dimensional Heisenberg model \cite{abrahams2011}. This feature is also shared by CuSC whereas Cu atoms carry spins. However, the superconducting gap structures of FeSCs and CuSCs may be different \cite{chen2008}. There is also intense debate on the existence of magnetic fluctuations, especially in Li(Na)FeAs \cite{borisheko2010, chen2009, platt2011, taylor2011, mengwang2011}. Some of these debates are concentrated on the electronic structure near the Fermi surface. 

The most important topic in HTSC is electron-pairing mechanism. More generally speaking, what mechanism supports superconductivity to survive at such a high temperature? Bardeen-Cooper-Schrieffer (BCS) theory \cite{bcs} stated that the electron-pairing mechanism in conventional superconductors is phonon-assisted thanks to the Fermi-liquid behaviour of electrons in metals making the attraction overcome repulsion near the Fermi surface. However, it is broadly believed that phonon alone cannot lead to HTSC. Nevertheless, this belief has never been proven correct carefully. Based on this belief, people would rather like to attribute the dominant pairing mechanism to the magnetism induced by the anti-ferromagnetic (AFM) interaction \cite{wang2011}. However, the co-existence of the nearest-neighbour and next-nearest-neighbour AFM exchange interaction in FeSCs may lead to a competition between N\'eel ordering and collinear ordering \cite{rotter2008, cruz2008, ma2009}. This might be a reason why it is difficult to see the anti-ferromagnetic spin fluctuation in some of FeSCs, e.g., LiFeAs. 

A proper description of magnetic properties and electronic structure is crucial for understanding of the mechanism of HTSC. However, most of the first-principles calculations performed so far within density functional theory (DFT) formalism are in the level of local density approximation (LDA) and generalized gradient approximation (GGA). The combination of LDA and dynamical mean field theory (DMFT) has been used to model the electronic structure  of a broad range of FeSCs very well \cite{yin2011, ferber2012}. On the other hand, the correlation effect can also be included by using hybrid exchange functional \cite{giovannetti2012}. The hybrid functional such as B3LYP \cite{b3lyp} and PBE0 \cite{adamo1998} can balance the tendencies of delocalising and localising wave functions and partially eliminates the self-interaction error by mixing exact exchange. For example, the band gap in various compounds can be predicted more accurately by hybrid functional than LDA or GGA\cite{muscat2001} and the effect of electron correlation in $\mathrm{CsC}_{60}$ can be assessed using hybrid exchange functional \cite{giovannetti2012}. A satisfactory theoretical description of electronic structure of FeSCs which can take electron correlation into account properly within DFT formalism is still rare \cite{platt2011, ma2009, singh2008, brydon2011,  subedi2008,  kusakabe2009, zhang2010, hozoi2009}. As the hybrid functional can balance the electron delocalization and localization, it is also possible to investigate the competition between superconductivity and spin fluctuations \cite{zhang2010b} by using hybrid functional. The computed electronic structure can be used to compare with that derived from experiments, e.g., angular resolved photoemission spectroscopy (ARPES), and the magnetic properties can be compared with the neutron scattering experiments. 

In this paper I report first-principles calculations for the electronic structure of LiFeAs and FeSe using DFT and hybrid functional PBE0 \cite{adamo1998}. The rest of the content is organized as follows: in \S\ref{sec:computationaldetails} I will introduce the computational details, in \S\ref{sec:resultsanddiscussions} I will show my calculation results and discuss them, and in \S\ref{sec:conclusion} I will draw some conclusions.  

\section{Computational details}\label{sec:computationaldetails}

The calculations for electronic structures of LiFeAs and FeSe are carried out by using DFT and hybrid functional PBE0 \cite{adamo1998} as implemented in the CRYSTAL 09 code \cite{crystal09}. The basis sets specially designed for Li\cite{merawa2004}, Fe\cite{moreira2000}, As\cite{curtiss1995}, Se\cite{crystalweb} atoms in solid-state compounds are used throughout all the calculations. The Monkhorst-Pack samplings \cite{packmonkhorst} of reciprocal space are carried out choosing a grid of shrinking factor to be $10\times 10 \times 6$ in order to be consistent with the ratios among reciprocal lattice parameters in LiFeAs and FeSe. Their experimental tetragonal structures \cite{shein2009, li2010} tabulated in Table \ref{tb:structure} are adopted in all the calculations. The truncation of the Coulomb and exchange series in direct space is controlled by setting the Gaussian overlap tolerance criteria to $10^{-7}, 10^{-7}, 10^{-7}, 10^{-7}$, and $10^{-14}$ \cite{crystal09}. The self-consistent field (SCF) procedure is converged to a tolerance of $10^{-8}$ a.u. per unit cell. To accelerate convergence of the SCF process, all calculations have been performed adopting a linear mixing of Fock matrices by $30\%$ .

\begin{table*}[htbp]
    \begin{tabular}{ | l | l | l | l |}
    \hline
     & Symmetry & Lattice constant ($\mathrm{\AA}$) & Atomic positions (in fraction) \\ \hline
    LiFeAs & P4/nmm (129) & a=b=3.791; c=6.364  & Li (0.25 0.25 0.83); Fe (0.75 0.25 0.5); As (0.25 0.25 0.29)  \\ \hline
    FeSe   & P4/nmm (129) & a=b=3.768;c=5.519 & Fe(0.25, -0.25, 0.0); Se(0.25, 0.25, 0.29)   \\ \hline
    \end{tabular}
    \caption{The structural parameters for LiFeAs and FeSe.}\label{tb:structure}
\end{table*}

Electronic exchange and correlation are described using the PBE0 hybrid functional \cite{adamo1998} which is free of empirical parameters. The advantages of PBE0 include a partial elimination of the self-interaction error and balancing the tendencies to delocalize and localize wave-functions by mixing a quarter of Fock exchange with that from a generalized gradient approximation (GGA) exchange functional \cite{adamo1998}. 

\begin{figure}[htbp]
\includegraphics[scale=0.4,clip=true]{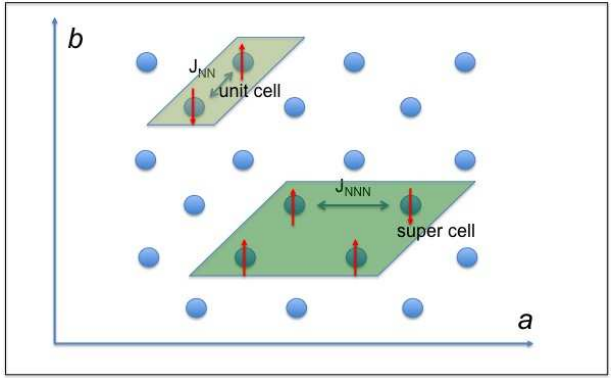}\\
\caption{(Colour on line.) A sketch for the two-dimensional Fe layer in LiFeAs and FeSe is shown. Fe atoms are labelled by blue balls. $J_{\mathrm{NN}}$ is the nearest neighbour exchange interaction, and $J_{\mathrm{NNN}}$ is the next-nearest neighbour exchange interaction. I also show the unit cell and super-cell (doubling the unit cell along lattice vector $a$) for the computation of $J_{\mathrm{NN}}$ and $J_{\mathrm{NNN}}$ respectively. The spin configurations in the unit cell and super-cell are used to calculate the total energies of AFM configuration $E_{\mathrm{AFM}}^{\mathrm{NN}}$ and $E_{\mathrm{AFM}}^{\mathrm{NNN}}$ respectively.}\label{fig:twod}
\end{figure}

The broken-symmetry method \cite{noodleman} is used to localize opposite electron spins on each atom in order to describe the AFM state. The exchange couplings in Heisenberg model \cite{heisenberg} is defined here as,
\begin{equation}\label{eq:spinh}
\hat{H}=J_{\mathrm{NN}}\sum_{ij\in \mathrm{NN}}{\hat{\vec{S}}_i\cdot\hat{\vec{S}}_j}+J_{\mathrm{NNN}}\sum_{ij\in \mathrm{NNN}}{\hat{\vec{S}}_{i}\cdot\hat{\vec{S}}_j},
\end{equation}
and determined by
\begin{eqnarray}\label{eq:DeltaE}
J_{\mathrm{NN}}&=&(E_{\mathrm{FM}}^{\mathrm{NN}}-E_{\mathrm{AFM}}^{\mathrm{NN}})/(8S^2),
\\J_{\mathrm{NNN}}&=&(E_{\mathrm{FM}}^{\mathrm{NNN}}-E_{\mathrm{AFM}}^{\mathrm{NNN}})/(4S^2)-2J_{\mathrm{NN}}.
\end{eqnarray}
where $E_{\mathrm{AFM}}^{\mathrm{NN}}$ and $E_{\mathrm{FM}}^{\mathrm{NN}}$ are defined as the total energies of a unit cell, in which the nearest neighbouring spins are in  anti-parallel (AFM configuration) or parallel (FM configuration), respectively. $J_{\mathrm{NN}}$ is defined as the nearest-neighbouring exchange interaction. The exchange interaction with the next nearest neighbour $J_{\mathrm{NNN}}$, is calculated using a $2a\times b\times c$ super-cell as shown in Fig.\ref{fig:twod}. $E_{\mathrm{AFM}}^{\mathrm{NNN}}$ is defined as the total energy for the state in which the next-nearest-neighbouring spin is flipped compared to FM configuration for a super-cell (Fig.\ref{fig:twod}). These formulae are obtained based on the two-dimensional lattice formed by Fe atoms as shown in Fig.\ref{fig:twod}. The total energies of stripe-type and checkerboard AFM ordering are computed by using the super-cell of which the new lattice vectors are defined as $\vec{a}^\prime=\vec{a}+\vec{b},\vec{b}^\prime=\vec{a}-\vec{b}$, and $\vec{c}^\prime=\vec{c}$ (the spin configurations are formed accordingly).

DFT total-energy calculations have some intrinsic disadvantages for computing exchange interactions, e.g., a comparison between two very large numbers to get a small exchange splitting in an all-electron local basis set formalism. However, the performance of the hybrid functional, e.g., B3LYP or PBE0 as implemented in CRYSTAL code \cite{crystal09} has previously been shown to provide an accurate description of the electronic structure and magnetic properties for both inorganic and organic compounds \cite{muscat2001, illas00, heutz2007, wu2008, wu2011}. 

The spin states of Fe atoms in LiFeAs and FeSe are still under active debate \cite{borisheko2010, chen2009, platt2011, taylor2011, mengwang2011}. In this paper, the total energies of the unit cell with non-magnetic and spin-2 state for Fe atoms are calculated for LiFeAs and FeSe carefully to determine the most stable one. 

\section{Results and discussions}\label{sec:resultsanddiscussions}
The calculations of a unit cell in LiFeAs and FeSe with $4$ unpaired electrons on Fe atoms (spin-2 state) are chosen to illustrate the electronic structure and magnetic properties.

\subsection{LiFeAs}\label{subsec:lifeas}
I first present the calculation results for electronic structure and magnetic properties of LiFeAs. For band structure, in total 21 bands are plotted for both AFM and FM configurations. 11 occupied bands are plotted for spin-up and spin-down in AFM configuration. 15(7) occupied bands are plotted for spin-up (spin-down) in FM configuration. The zero of energy is chosen to be the Fermi energy. In density of states (DOS) the spin-up is positive and spin-down negative. The projected DOS (PDOS) to $2s$ and $2p$-orbitals of Li is in red, to $4s$ and $4p$ of Fe in green, $d$ of of Fe in blue, and $4s$ and $4p$ of As in cyan. 

\begin{figure*}[htbp]
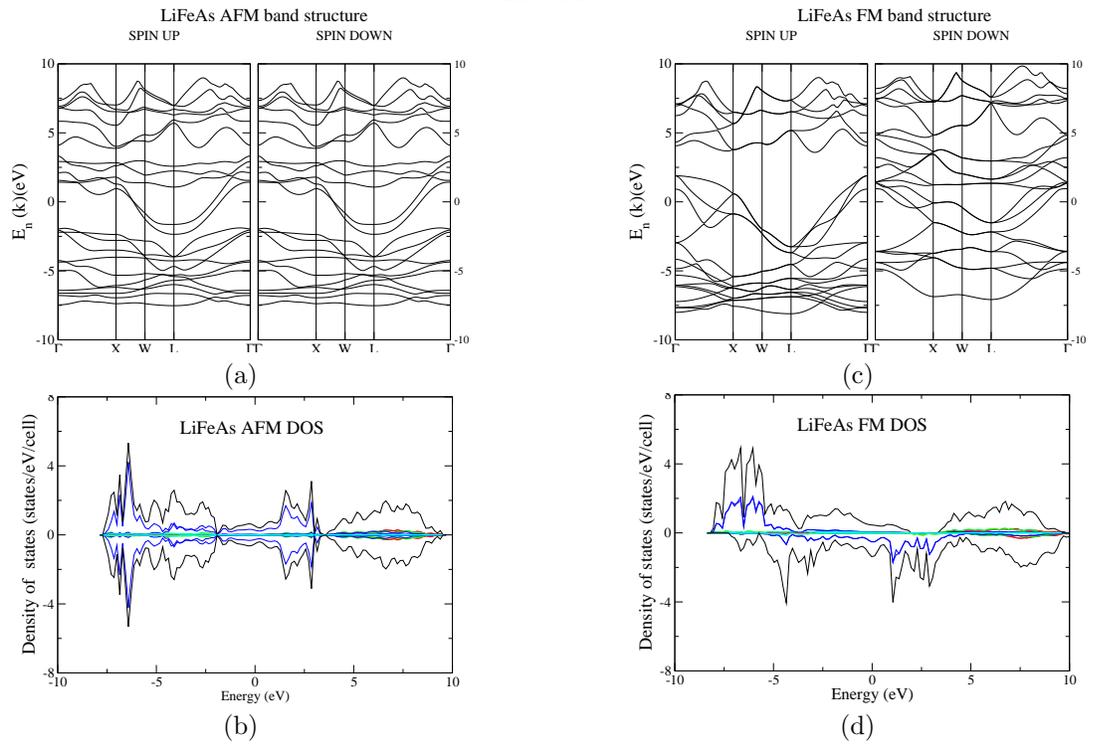

\begin{tabular}{cccc}
\centering
 & &\large{LiFeAs} & \\
&\includegraphics[scale=0.25,clip=true,trim=0mm 1mm 0mm 2mm]{LiFeAs_afm_band.eps}&&\includegraphics[scale=0.25,clip=true, trim=0mm 1mm 0mm 2mm]{LiFeAs_fm_band.eps}\\
&(a)&&(c)\\
&\includegraphics[scale=0.25,clip=true,trim=0mm 1mm 0mm 2mm]{LiFeAs_afm_dos.eps}&&\includegraphics[scale=0.25,clip=true, trim=0mm 1mm 0mm 2mm]{LiFeAs_fm_dos.eps}\\
&(b)&&(d)\\
\end{tabular}
\caption{ (Colour on line.) The band structure and density of states of LiFeAs in AFM (a,b) and FM (c,d) configurations are shown. The projected density of states to $2s$ and $2p$ of Li is in red, $4sp$ of Fe in green, $3d$ of Fe in blue, $4s$ and $4p$ of As in cyan, and the total one in black. }\label{lifeas_afmfm}
\end{figure*}

In Fig.\ref{lifeas_afmfm} (a) and (b) the band structure and density of states (DOS) are shown when the spins on Fe atoms in a unit cell are anti-parallel (AFM configuration). As shown in PDOS the d-bands is dominant at the Fermi surface. In Fig.\ref{lifeas_afmfm} (c) and (d), the band structure and density of states (DOS) are presented when the spins on Fe atoms in a unit cell are parallel (FM configuration). The computed band structure suggests that LiFeAs is metallic. This is consistent with the previous experimental and theoretical results \cite{yin2011}. The Mulliken charge on Li is $\sim +0.7 |e|$, Fe $\sim + 0.5 |e|$, and As $\sim -1.2 |e|$ for AFM and FM configurations. The Mulliken spin densities on Fe atoms with spin anti-parallel are $\sim  \pm 3.1\mu_B$ for AFM configuration. The discrepancy between this value and the expect one, i.e., $4 \mu_B$ may be due to the hybridization between Fe and As and the closer distance between Fe atoms compared to other FeScs \cite{yin2011}. For FM configuration, the Mulliken spin densities on Fe atoms are $\sim 3.2 \mu_B$ and on As are $\sim 0.2 \mu_B$. This indicates there is a stronger hybridization between Fe and As in FM than AFM configuration, leading to the super-exchange mechanism between Fe atoms. In addition, this value of Mulliken spin densities on Fe atoms is larger than those computed ($\sim 0.2 \mu_B$) by using PBE \cite{pbe} functional. It is interesting that there is a large magnetic moment on Fe atoms although the band structure as whole shows that LiFeAs is metallic.

The computed total energies per unit cell for the non-magnetic state is higher than that of the spin-2 state by $2.45$ eV. This energy difference is in a good agreement with that found in Ref.\cite{hozoi2009} (2.34 eV), where the post-Hartree-Fock quantum chemistry methods including multiconfiguration completeactive-space self-consistent-field (CASSCF) and multireference configuration-interaction (MRCI) have been used to determine the ground state. The exchange interaction in LiFeAs $J_{\mathrm{NN}}$ is $\sim 14$ meV and $J_{\mathrm{NNN}}$ is $\sim 17$ meV. The result that $J_{\mathrm{NNN}}$ is larger than $J_{\mathrm{NN}}$ might be due to different bond angles for the nearest neighbour and next nearest neighbour. The energy of the stripe-type AFM ordering is also computed, which is higher than that of the checkerboard one by $\sim 0.8$ eV per cell. 

The prediction of an AFM ground state in this work is qualitatively in agreement with recent theoretical results in Ref.\cite{singh2008}.  In Ref.\cite{singh2008}, it was also concluded that there is an AFM ground state in LiFeAs. However, the details of the band structure in Ref.\cite{singh2008}, which largely depends on the functional, are different from the work presented here. The previous work in Ref.\cite{brydon2011} suggests that the absence of AFM ordering and points to a triplet pairing mechanism. Based on my results, the system will prefer a collinear ordering because $J_{\mathrm{NNN}}> J_{\mathrm{NN}}/2$, i.e., each spin has two ferromagnetically coupling neighbours and another two anti-ferromagnetically coupled ones. Therefore my work is partially consistent with the work in Ref.\cite{brydon2011}. However, my work disagrees with Ref.\cite{zhang2010} because they predicted that AFM configuration has a higher energy than the non-magnetic one. This might be due to the exclusion of the exact exchange in the functional used in \cite{zhang2010}. My work is consistent with the experiments that have shown that there may be strong anti-ferromagnetic fluctuations \cite{taylor2011,mengwang2011}.

\subsection{FeSe}
\begin{figure*}[htbp]
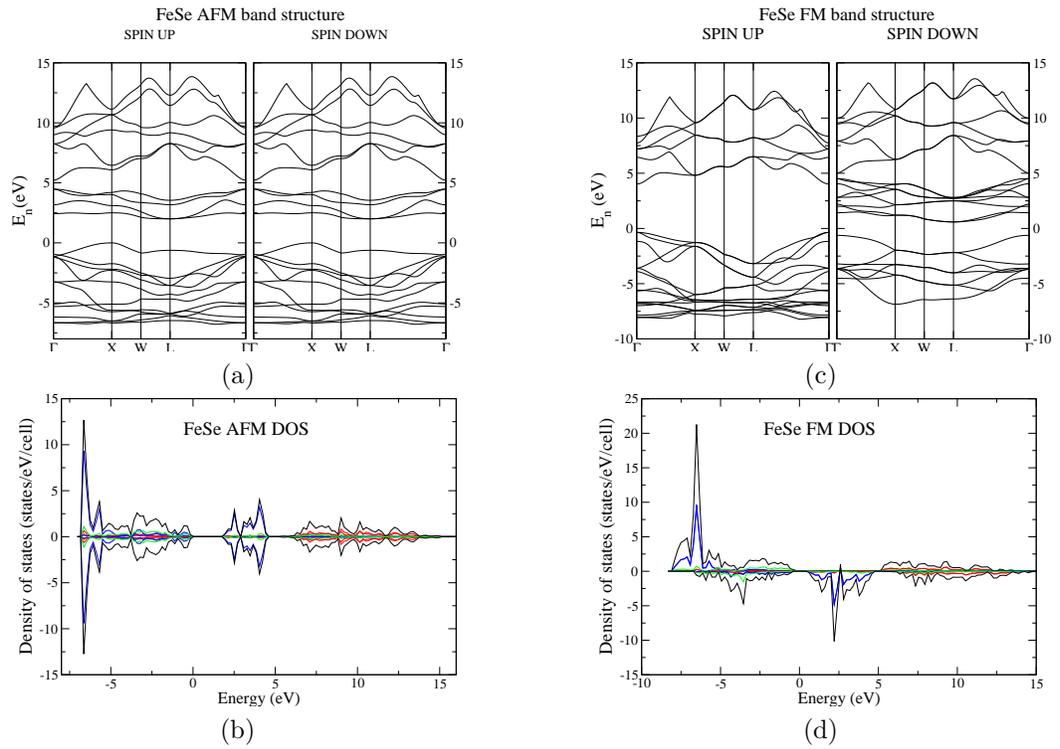

\begin{tabular}{cccc}
\centering
 & &\large{FeSe} & \\
&\includegraphics[scale=0.25,clip=true,trim=0mm 1mm 0mm 2mm]{FeSe_afm_band.eps}&&\includegraphics[scale=0.25,clip=true,trim=0mm 1mm 0mm 2mm]{FeSe_fm_band.eps}\\
&(a)&&(c)\\
&\includegraphics[scale=0.25,clip=true,trim=0mm 1mm 0mm 0mm]{FeSe_afm_dos.eps}&&\includegraphics[scale=0.25,clip=true,trim=0mm 1mm 0mm 1mm]{FeSe_fm_dos.eps}\\
&(b)&&(d)\\
\end{tabular}
\caption{(Colour on line.) The band structure and density of states of FeSe AFM (a, b) and FM (c, d) configurations respectively are shown. Notice that the band structure suggests that AFM and FM configurations are insulating. The PDOS of $4sp$ orbitals of Se is in red, $3d$ of Fe in blue, $4sp$ of Fe in green, and the total one in black.}\label{fese_afmfm}
\end{figure*}

The band structure and DOS for AFM (FM) configuration of FeSe are shown in Fig.\ref{fese_afmfm} (a,b), and  Fig.\ref{fese_afmfm} (c,d), respectively. In total 21 bands are plotted for both AFM and FM configurations. 11 occupied bands are plotted for spin-up and spin-down in AFM configuration. 15(7) occupied bands are plotted for spin-up (spin-down) in FM configuration. The zero of energy is chosen to be the Fermi energy. In density of states (DOS) the spin-up is positive and spin-down negative. The projected DOS (PDOS) to $4sp$ of Fe is in red, $d$ of of Fe in blue, and $4sp$ of Se in green. 

As shown in the band structures of FeSe (Fig.\ref{fese_afmfm}) both AFM and FM configurations are insulating although they have different band gap. For AFM the band gap is $\sim 2.0$ eV whereas for FM the band gap is $\sim 5.0$ eV for spin-up and $\sim 2.0$ eV for spin-down. This is qualitatively different from that in LiFeAs. Once again, the $d$-orbitals of Fe atoms have the dominant contribution to the bands near the Fermi energy. The Mulliken charge on Fe is $\sim +0.3 |e|$ and on Se $\sim -0.3 |e|$ for both AFM and FM configurations. The Mulliken spin densities for Fe atoms are $\pm 3.6 \mu_B$ in the AFM configuration. This value is close to the expected, i.e., $4 \mu_B$. For FM configuration, the Mulliken spin densities for Fe atoms are $3.7 \mu_B$ and for Se $0.3 \mu_B$. This also indicates the hybridization between Fe and Se atoms. In addition, the calculation using PBE functional shows that FeSe is metallic. This suggests that the corrected functional PBE0 can improve the exchange-correlation functional and hence the description of electronic structure. 

I also compare the total energies per unit cell of FeSe for different spin configurations including non-magnetic and spin-2. The total energies per cell of the non-magnetic state higher than that of spin-2 by $5.80$ eV. This indicates that the spin-2 state is also favoured by FeSe in this tetragonal structure. For FeSe, $\mathrm{J}_\mathrm{NN}$ is approximately $6$ meVs and $\mathrm{J}_\mathrm{NNN}$ is approximately $13$ meVs. This is qualitatively in agreement with the results for FeSe in Ref.\cite{ma2009}. The strong hybridization between Fe and Se in FM configuration indicate the exchange mechanism is dominated by super-exchange mediated by Se atoms. The energy of the stripe-type AFM ordering is also computed, which is lower than that of the checkerboard one by $\sim 0.1$ eV per cell. 

\section{Summary and conclusions} \label{sec:conclusion}

The total energies of different spin configurations including non-magnetic and spin-2 have been computed by using hybrid functional PBE0 and are carefully compared for LiFeAs and FeSe, which shows that both of them favour spin-2 state.

The band structure calculations show that LiFeAs is conducting, but FeSe is insulating. The fundamental difference between the band structures of LiFeAs and FeSe might be able to shed some light on the comparison between the experiments on them. As shown in the electronic structure of LiFeAs, the co-existence of localized magnetic moments and the dominant $d$-band near the Fermi surface might lead to the competition between itinerant and localized $d$-orbital magnetism, and this might be the key to understand the superconductivity of LiFeAs \cite{dai2012}. 

 The exchange interactions of the nearest and next nearest neighbours are also calculated, and the computed values qualitatively agree with recently experiments and theoretical results leading to a two-dimensional collinear ordering rather than N\'eel ordering. The energies of checkerboard and stripe-type AFM ordering have been computed. For LiFeAs the checkerboard one is lower whereas for FeSe the stripe-type is lower. 

These calculations might be useful for the experimentalists who are concerned with the superconductivity, magnetic properties, and electronic structure of the 111-type and 11-type FeSCs. In the future, I would like to investigate further how the localised and itinerant magnetisms compete with each other and the change of electronic structure in doped LiFeAs and FeSe to understand superconductivity in FeSCs.

\end{document}